# Waiting times between orders and trades in double-auction markets

Enrico Scalas[a,b], Taisei Kaizoji[c], Michael Kirchler[d], Jürgen Huber[d], Alessandra Tedeschi[e]

[a]Università del Piemonte Orientale, Italy
[b]INFM Genova, Italy
[c]International Christian University Tokyo, Japan
[d]Universität Innsbruck, Austria
[e]Università di Roma, Italy

**Abstract**

In this paper the survival function of waiting times between orders and the corresponding trades in a double-auction market is studied both by means of experiments and of empirical data. It turns out that, already at the level of order durations, the survival function cannot be represented by a single exponential, thus ruling out the hypothesis of constant activity during trading. This fact has direct consequences for market microstructural models. They must include such a non-exponential behaviour to be realistic.

**1. Introduction**

In recent times, the statistical properties of high-frequency financial data have been related to market microstructure both by means of empirical studies and agent-based models [1-7]. High-frequency econometrics is now well established after research on conditional duration models [8-11], compound Poisson processes [12-14] and continous-time random walks [15-26]

In high-frequency financial data not only returns but also waiting times between consecutive trades are random variables [27]. Already Lo and McKinlay [28] made this remark, but it can be traced at least to papers on the application of compound Poisson processes [29] and subordinated stochastic processes to finance [30,31]. This is a consequence of the asynchronous character of trading in financial markets. Many regulated markets implement trading via the so-called *continuous double-auction* where buyers and sellers send their orders at random times. These orders, including price and quantity (*volume*), are collected in a *book*. Depending on market regulations, traders may submit various kinds of orders. The most common are *limit orders*: orders to buy a specified quantity of a security at or below a given price, or to sell it at or above a given price (the *limit price*) and *market orders*: orders to buy or sell a specified quantity of a security at the best available price. When the *best bid* (buy order) and the *best ask* (sell order) match, a transaction (*trade*) takes place, securities are transferred from the seller to the buyer and money from the buyer to the seller.

It turns out that compound Poisson processes are an instance of continuous-time random walks (CTRWs) [32]. The application of CTRWs to problems in economics dates back, at least, to the 1980s. In 1984, Rudolf Hilfer discussed the application of stochastic processes to operational planning, and used CTRWs as tools for sale forecasts [33]. Already in 1903, the PhD thesis of Filip

Lundberg presented a model for ruin theory of insurance companies, later elaborated by Cramér [34,35]. The stochastic process for claims is another example of compound Poisson process and thus of CTRW.

As for the unconditional distribution of intertrade waiting times, *a priori*, there is no reason for independent market investors to place buy and sell orders in a time-correlated way. This argument would lead us to expect an exponential process for inter-order waiting times. Moreover, if the order selection were a simple thinning of the order process, then exponential waiting times should be expected between consecutive trades as well [36]. However, it turns out that the distribution of intertrade waiting times does not follow the exponential law. This has been shown for the 30 DJIA stocks traded at NYSE in October 1999 [17,23,24], a remark also made by Engle and Russel [8]. Results of many other independent research groups corroborate the finding that intertrade durations are non-exponentially distributed. Plerou et al. noticed an anomalous behaviour of the number of transactions over a given time interval [37]. A study on the waiting times in a contemporary FOREX exchange and in the XIXth century Irish stock market was presented by Sabatelli et al. [38]. They were able to fit the Irish data by means of a Mittag-Leffler function as Mainardi et al. did before in a paper on the waiting-time marginal distribution in the German Bund-future market [16]. Kim and Yoon studied the tick dynamical behavior of the bond futures in Korean Futures Exchange (KOFEX) market and found that the survival probability displays a stretched-exponential form [39]. This finding is also present in an empirical study of NYSE data [40]. Moreover, just to stress the relevance of non-exponential waiting times, a power-law distribution has been recently detected by T. Kaizoji and M. Kaizoji in analyzing the calm time interval of price changes in the Japanese market [41].

In this paper, we present results on the order selection process for a double auction for both well-controlled experiments and for real market data. The main result of this paper is that also waiting-times between consecutive orders do not follow the exponential distribution.

The paper is organized as follows. Section 2 is devoted to the experimental results. In Section 3, the empirical analysis on two LSE stocks traded in 2002 is presented. Finally, in Section 4, the main conclusions are discussed.

**2. Experimental results**

*2.1 Method*

The experiments have been performed at the University of Innsbruck by two of us (Jürgen Huber and Michael Kirchler). They were based on an asymmetric information system. Nine traders (students of Economics at the University of Innsbruck) with different forecasting abilities trade for 30 periods in a market that is a continuous double-auction. With 100 seconds for each period each session lasts 50 minutes. All parameters (especially interest rates and dividends) are set to let one period in the experiment correspond to one month in a real market. Each trader starts the experiment with 40 shares and 1,600 units of cash. This allows traders to buy or sell shares depending on their expectations about the future development of dividends. In the double auction market, traders can freely sell or buy shares by placing limit orders or market orders by accepting open bids and asks from other traders. However, short selling was not allowed.

The information structure of the market is such that agent (trader) $Ij$ knows dividends for $j$ periods, that is, agent $I1$ knows only the current dividend, agent $I2$ the dividends of this and the next period, and so on, up to agent $I9$ who is the best informed trader. Therefore, in this market there are asymmetrically informed traders who are able to predict future cash flows to a different extent. This information structure is common knowledge.

At the start of each period, subjects get information on future dividends according to their information level. In addition we display to each trader the net present value of the stock given his/her information. This is derived using Gordon's formula, discounting the known dividends and using the last one as a continuous, infinite stream which is also discounted.

$$E(V, I_{j,k}) = \frac{D_{k+j-1}}{(1+r_e)^{j-2} \cdot r_e} + \sum_{i=k}^{k+j-2} \frac{D_i}{(1+r_e)^{i-k}} \quad (1)$$

$E(V, I_{j,k})$ stands for the conditional expected value of the asset in period $k$, $j$ represents the index for the information level of the traders, and $r_e$ is the risk adjusted interest rate in period $k$. As we can see from equation (1), the dividend in $(k+j-1)$, namely the last dividend known to trader $j$ is assumed to remain constant for an infinite number of periods. All the other dividends are also discounted with $r_e$. The resulting paths of the conditional expected values of the asset for odd information levels with $k=30$ periods in our experimental treatment are shown in Figure 1.

[Fig1 here]

*Figure 1. Conditional expected values as a function of period (for visibility we only show the odd information levels)*

Beginning with *I9* the functions in Figure 1 are shifted for each information level *Ij* by *(9-j)* periods to the right, showing a main characteristic of our model, namely that better informed traders receive information earlier than less informed ones. So, information on the intrinsic value of the company that trader *I9* sees in one period is seen by trader *I8* one period later, and by trader *I1* eight periods later, giving the better informed an informational advantage. For more details on the design, see reference [42].

*2.1 Results*

Six experiments have been performed so far. The results for the survival probability function:

$$\Psi(\tau) = \Pr\{t > \tau\} \quad (2)$$

are given in Figure 2, both for orders and trades, and compared with the exponential distribution. A visual inspection shows that at least in five of six cases, the measured data do not follow the exponential distribution.

[Fig2 here]

*Figure 2: Waiting-time survival functions of orders (dots) and trades (crosses) in seconds for the experimental markets. The solid line represent the standard exponential survival function for orders and the dash-dotted line for trades.*

As we have argued above, in the presence of constant activity in the market, the survival function is an exponential function:

$$\Psi(\tau) = \exp(-\tau/\tau_0), \quad (3)$$

where $\tau_0$ is the average duration. It can be shown that, if the waiting times between two consecutive events follow the exponential distribution then the number $n$ of events in a given time interval is a random variable characterized by the Poisson distribution.

Indeed, the probability $P(n,t)$ of having $n$ events from time $t_0 = 0$ up to time $t$ is given by the following convolution:

$$P(n,t) = \int_0^t \int_0^{t_n} \cdots \int_0^{t_1} dt_1 \cdots dt_n \, \Psi(t-t_n) \psi(t_n - t_{n-1}) \cdots \psi(t_2 - t_1) \psi(t_1 - t_0). \tag{4}$$

where $\psi(\tau)$ is the probability density of waiting times. The Laplace transform of $P(n,t)$ is:

$$\tilde{P}(n,s) = \tilde{\Psi}(s)[\tilde{\psi}(s)]^n. \tag{5}$$

Recalling that:

$$\psi(\tau) = -\frac{d\Psi(\tau)}{d\tau} = \frac{1}{\tau_0} \exp(-\tau/\tau_0), \tag{6}$$

combining equation (3) and (5), and inverting the Laplace transform, one gets:

$$P(n,t) = \frac{(t/\tau_0)^n}{n!} \exp(-t/\tau_0) \tag{7}$$

Moreover, the exponential distribution is the only *memoryless* distribution for a point process (Cox, 1979). Therefore, it is natural to test the empirical survival function against the above exponential model. A suitable statistical test is the Anderson-Darling (AD) test [43]. In order to perform the test, the statistics $A^2$ is computed according to the following rule:

$$A^2 = \left(-\sum_{i=1}^n \frac{(2i-1)}{n}[\ln \Psi(\tau_{n+1-i}) + \ln(1 - \Psi(\tau_i))] - n\right) \times (1 + (0.6/n)) \tag{8}$$

where $\tau_1 \leq \tau_2 \leq \ldots \leq \tau_n$.

In Table 1, the values of the AD $A^2$ statistics are given for all the six experimental markets. In all these cases the null hypothesis of exponentially distributed waiting times can be rejected at the 1 % significance level (the limiting value is 1.957). These values are compared with those that can be obtained from exponentially distributed waiting times. The results of experimental market 3 are closer to the exponential.

**Table 1.** AD statistics $A^2$ [43] for the investigated data sets (O: orders; T: trades). $N$ is the number of non-zero samples. For comparison, $A^2_{th}$ is the corresponding value of the AD statistics for an equivalent number of exponentially distributed waiting times. The fifth and sixth columns contain the average waiting time and the standard deviation of the data sets, respectively.

| Market   | $N$  | $A^2$  | $A^2_{th}$ | Average w.t | σ w.t. |
|----------|------|--------|------------|-------------|--------|
| Exp. 1 O | 1240 | 104.4  | 0.46       | 1.9 s       | 2.1 s  |
| Exp. 1 T | 746  | 28.2   | 0.56       | 3.3 s       | 4.1 s  |
| Exp. 2 O | 1113 | 79.4   | 0.48       | 2.1 s       | 2.2 s  |
| Exp. 2 T | 478  | 10.9   | 0.58       | 5.3 s       | 7.0 s  |
| Exp. 3 O | 862  | 40.7   | 0.95       | 3.0 s       | 3.0 s  |
| Exp. 3 T | 280  | 2.6    | 0.84       | 9.6 s       | 9.9 s  |
| Exp. 4 O | 1465 | 203.7  | 1.01       | 1.3 s       | 1.4 s  |
| Exp. 4 T | 836  | 45.6   | 1.36       | 2.8 s       | 3.3 s  |
| Exp. 5 O | 1344 | 158.0  | 0.36       | 1.5 s       | 1.6 s  |
| Exp. 5 T | 618  | 28.1   | 0.37       | 3.5 s       | 5.5 s  |
| Exp. 6 O | 1446 | 194.1  | 0.68       | 1.3 s       | 1.4 s  |
| Exp. 6 T | 830  | 45.0   | 0.17       | 2.6 s       | 3.4 s  |

## 3. Empirical results on the survival function

*3.1 The LSE data*

The waiting-time data have been extracted from the historical database of the London Stock Exchange where orders and quotes are stored for the electronic market; these data are a significant fraction, but do not include all the orders and quotes. Our data set consists of waiting times between orders and trades for both Glaxo Smith Kline (GSK) and Vodafone (VOD) stocks traded in the following months: March, June, and October 2002. Nearly 800,000 orders and 540,000 trades have been analyzed. Both limit and market orders have been included. For orders and trades of each stock, the average waiting times and the standard deviations are given in Table 2 below. The difference between these two values already points to a non-exponential distribution of durations. The use of one-month high-frequency data is a trade-off between the necessity of managing enough data for significant statistical analyses and, on the other hand, the goal of minimizing the effect of external economic fluctuations.

*3.2 The Anderson-Darling test on the survival function*

In Table 2, the values of the AD $A^2$ statistics are given for all the six monthly data sets (three for each company). Again, in all these cases the null hypothesis of exponentially distributed waiting times can be rejected at the 1 % significance level (the limiting value is 1.957). It is therefore safe to conclude that the survival function for waiting times for both orders and trades is non-exponential. The last two columns of Table 2 show that there is significant excess standard deviation in the distribution of waiting times, a result that corroborates the rejection of the null hypothesis.

**Table 2. AD statistics $A^2$ (Stephens, 1974) for the investigated data sets (O: orders; T: trades). The third column gives the larger waiting time taken into account to compute the $A^2$ statistics and avoid overflows. The fourth and fifth columns contain the average waiting time and the standard deviation of the raw truncated data sets, respectively.**

| Stock | N | $A^2$ | Larger w.t. | Average w.t | σ w.t. |
|---|---|---|---|---|---|
| VOD-03 O | 92887 | 1642 | 300 s | 6.8 s | 10.7 s |
| VOD-03 T | 82721 | 3273 | 300 s | 7.9 s | 14.5 s |
| VOD-05 O | 110588 | 2232 | 200 s | 6.2 s | 10.2 s |
| VOD-05 T | 96132 | 3594 | 200 s | 7.5 s | 14.1 s |
| VOD-10 O | 176940 | 5387 | 200 s | 4.1 s | 6.4 s |
| VOD-10 T | 105787 | 4531 | 200 s | 7.2 s | 13.7 s |
| GSK-03 O | 98882 | 2201 | 200 s | 6.4 s | 10.7 s |
| GSK-03 T | 67068 | 1826 | 200 s | 9.5 s | 15.9 s |
| GSK-05 O | 93237 | 2083 | 200 s | 7.4 s | 12.2 s |
| GSK-05 T | 67263 | 1913 | 200 s | 10.5 s | 18.6 s |
| GSK-10 O | 157811 | 4230 | 150 s | 4.6 s | 7.2 s |
| GSK-10 T | 85272 | 2684 | 150 s | 8.5 s | 15.0 s |

For comparison with the experimental results, in Figure 3, we plot the empirical survival function for GSK orders and trades in March 2002. Also in this case, visual inspection already indicates that the distribution of waiting times is far from exponential.

[Fig3 here]

*Figure 3: Waiting-time survival functions for orders (dots) and trades (crosses) in seconds for GSK, March 2002. The solid line represent the standard exponential survival function for orders and the dash-dotted line for trades.*

### 4. Discussion and conclusions

In this paper, we have shown that not only waiting times between consecutive trades are non-exponentially distributed, but the anomalous behaviour is also shared by waiting times between consecutive orders. This property is not only present in real market data, such as the LSE data studied here, but also in market experiments as shown in Section 2. The experiments demonstrate that variable activity during trading is present also on quite short time scales (each experimental period lasts 100 seconds).

This has direct consequences for market microstructural models. To be realistic, such models must include a variable activity yielding a non-exponential behaviour. Double-auction market simulations with constant activity [44,45] fail to take into account this feature and might miss subtle but important effects related to non-constant trader activity. Including the anomaly in a double-auction market mechanism is easy and leads to promising and interesting results [46]. Muchnik and Solomon have developed a computing framework which naturally takes into account the asynchronous character of trading in a financial market [47].

Recently, Barabási has argued that the timing of many human activities is described by non-Poisson statistics, characterized by bursts of rapidly occurring events followed by more relaxed periods [48]. In this paper we have provided clear evidence that this is precisely the case in a financial market, already at the level of the order process in a continuous double-auction.

Scalas et al. have provided a simple phenomenological way of taking into account variable human activity in a financial market in terms of exponential mixtures [23]. Suppose that the trading period can be divided into $N$ intervals of constant activity and characterized by average durations $\tau_{01}$, ..., $\tau_{0N}$. Then the survival function can be written as a weighted sum of exponentials:

$$\Psi(\tau) = \sum_{i=1}^{N} a_i \exp(-\tau/\tau_{0i}) \tag{9}$$

where $\sum_i a_i = 1$ so that $\Psi(0) = 1$. In general, the survival function given in equation (9) exhibits a significant deviation from the simple exponential behaviour for intermediate waiting times and can mimic the presence of heavy tails in the distribution of waiting times. For instance, in the simple case where the survival function is given by:

$$\Psi(\tau) = a_1 \exp(-\tau/\tau_{01}) + a_2 \exp(-\tau/\tau_{02}), \tag{10}$$

it is straightforward to show that the standard deviation is greater than the average value of the waiting times [49]. In fact, one has:

$$\langle \tau \rangle = a_1 \tau_{01} + a_2 \tau_{02}, \tag{11}$$

and

$$\langle \tau^2 \rangle = 2a_1 \tau_{01}^2 + 2a_2 \tau_{02}^2, \tag{12}$$

hence:

$$\sigma_\tau^2 - \langle \tau \rangle^2 = \langle \tau^2 \rangle - 2\langle \tau \rangle^2 = 2a_1 a_2 (\tau_{01} - \tau_{02})^2 > 0. \tag{13}$$

However, it might be impossible to find intervals of nearly constant activity in financial trading. In this case, the continuous version of equation (9) may still prove useful:

$$\Psi(\tau) = \int_0^\infty d\tau_0 \, g(\tau_0) \exp(-\tau/\tau_0) \tag{12}$$

where $g(\tau_0)$ is a continuous average duration spectrum subject to the condition

$$\int_0^\infty d\tau_0 \, g(\tau_0) = 1. \tag{13}$$

Indeed, many interesting waiting-time distributions can be written in the form specified by equation (10). Conversely, the numerical inversion of equation (10), a Fredholm equation of the first kind for the function $g(\tau_0)$, is possible from the empirical estimates of $\Psi(\tau)$.

Another phenomenological method of describing variable human activity in a financial market has been introduced by Bertram [50]. He has developed a threshold model, called threshold Brownian motion, where geometric Brownian motion is subordinated to a stochastic activity process with threshold.

The time has now come to further study asynchronous trading in a double-auction market by means of agent based models and this will be the subject of future research.

**Acknowledgements**

This work was supported by the Italian M.I.U.R. F.I.S.R. Project "High frequency dynamics of financial markets". Financial support from the Raiffeisen-Landesbank Tirol and the Center for Experimental Economics at the University of Innsbruck for the experiments is gratefully acknowledged

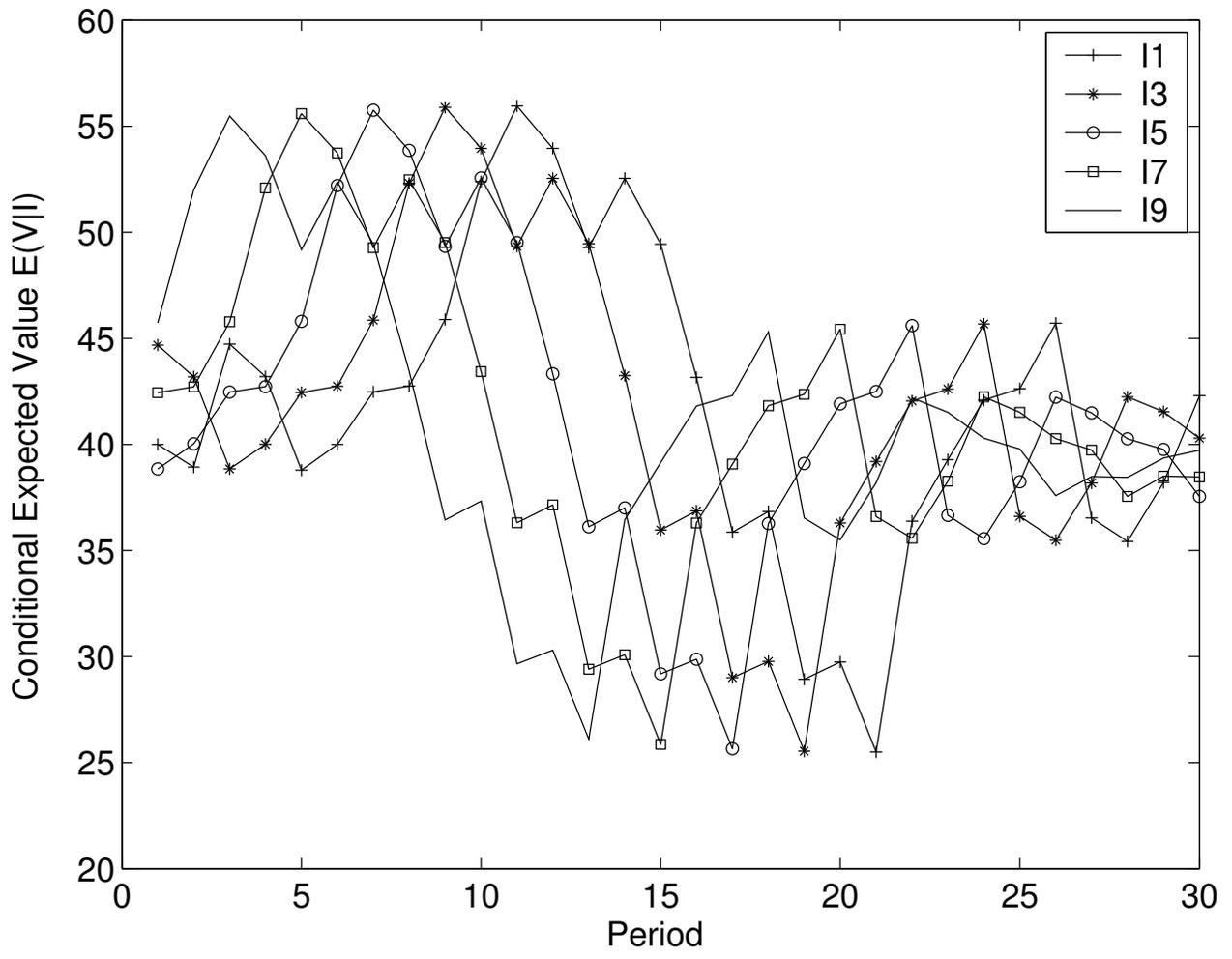

**Fig. 1**

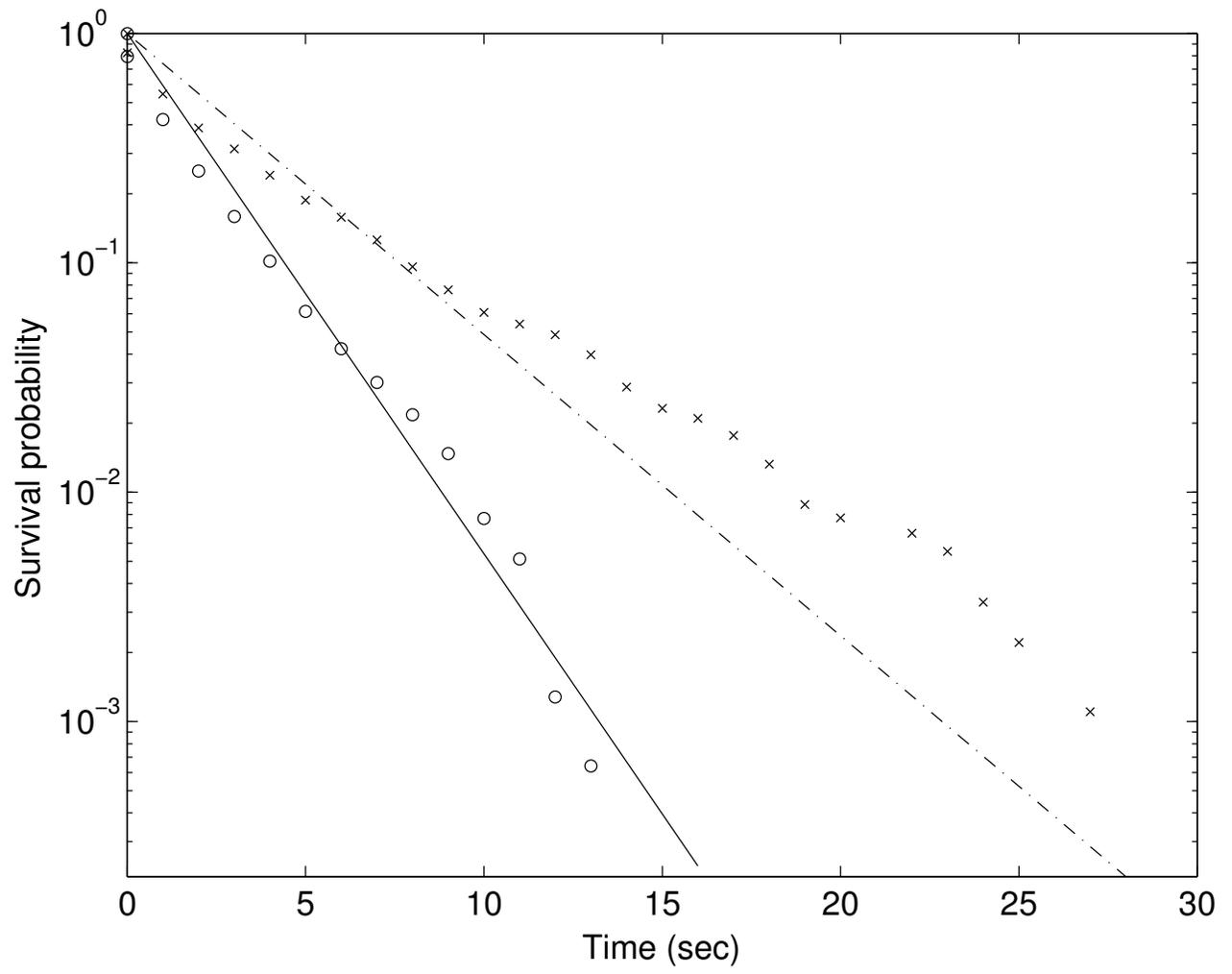

**Fig. 2a**

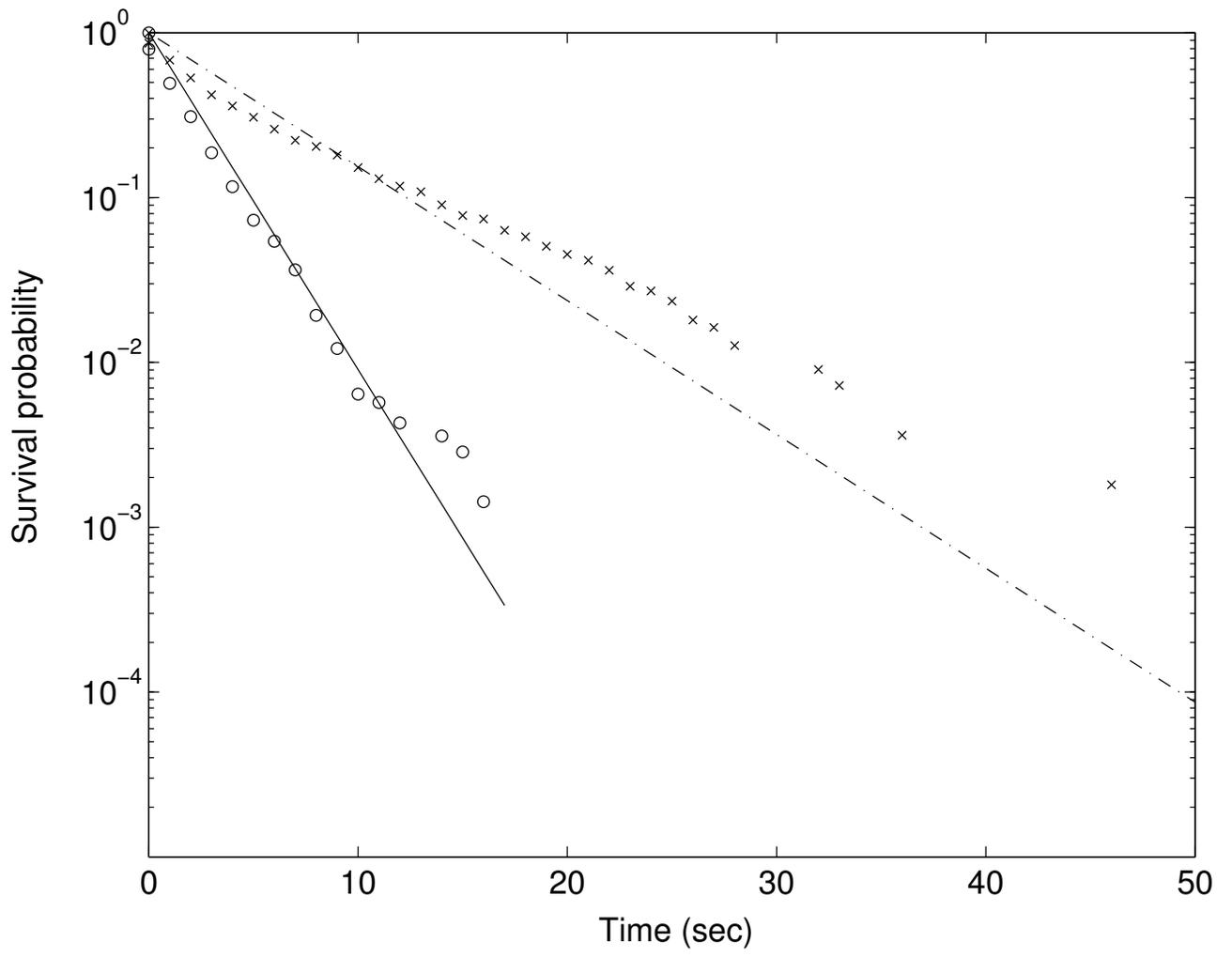

**Fig. 2b**

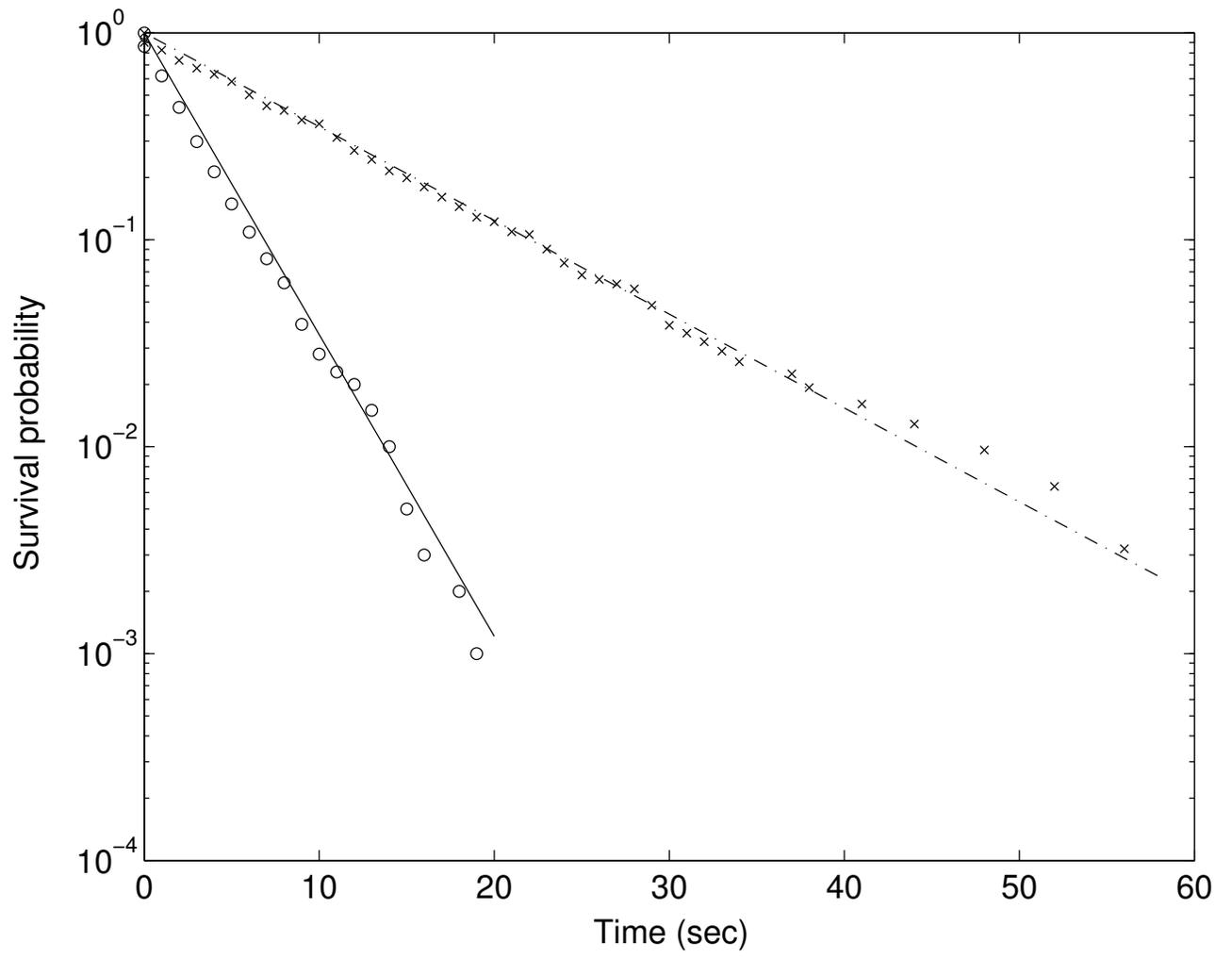

Fig. 2c

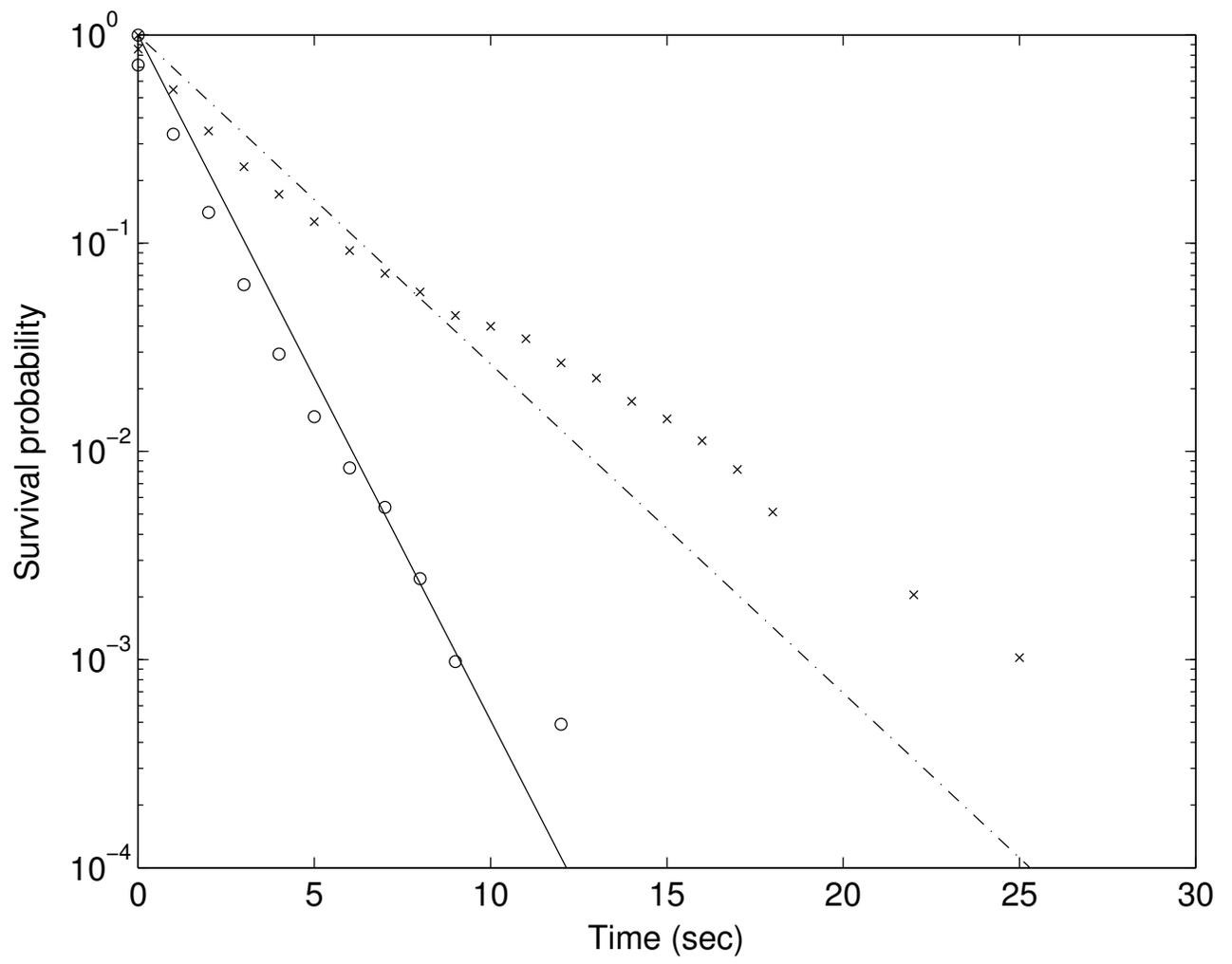

**Fig. 2d**

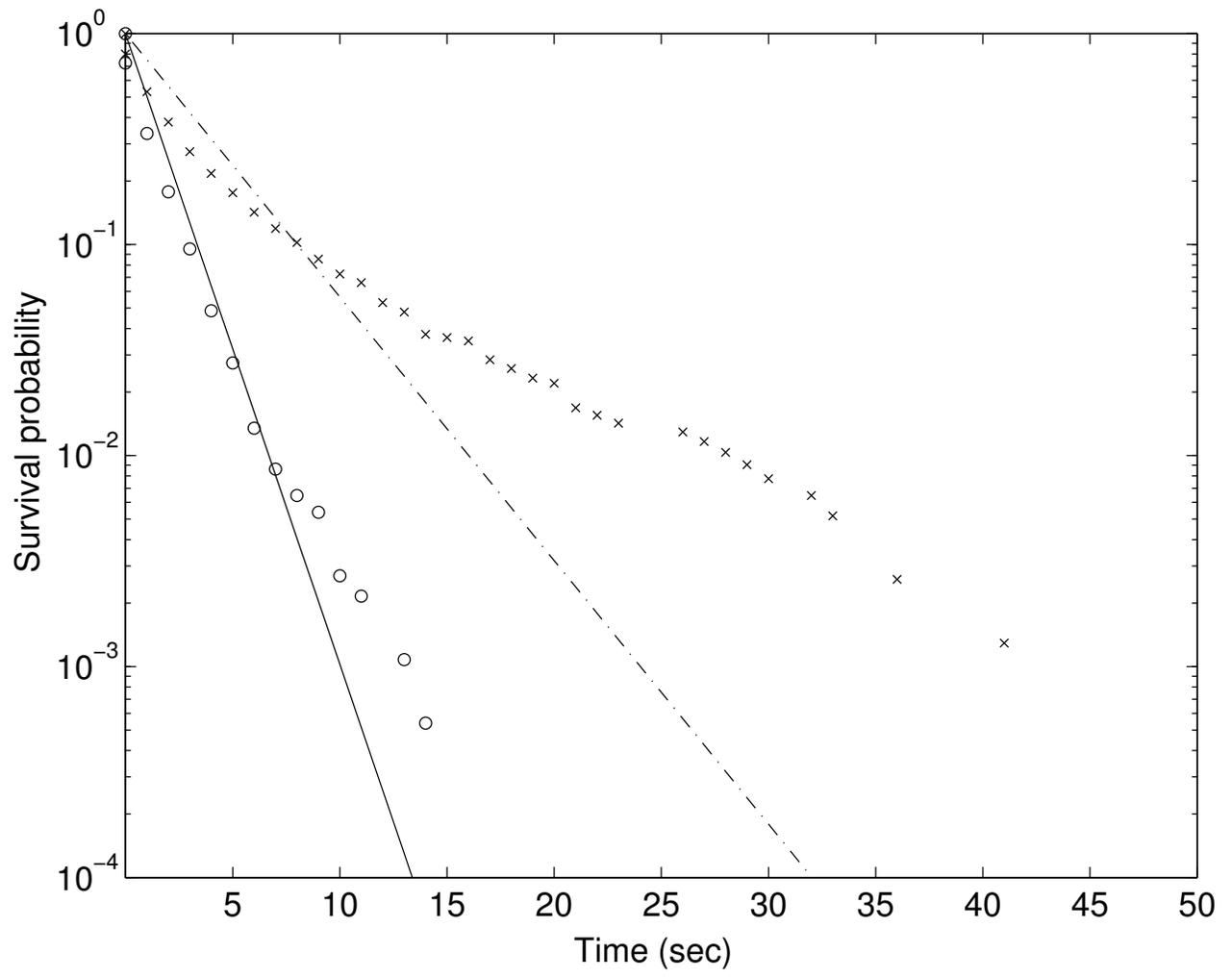

**Fig. 2e**

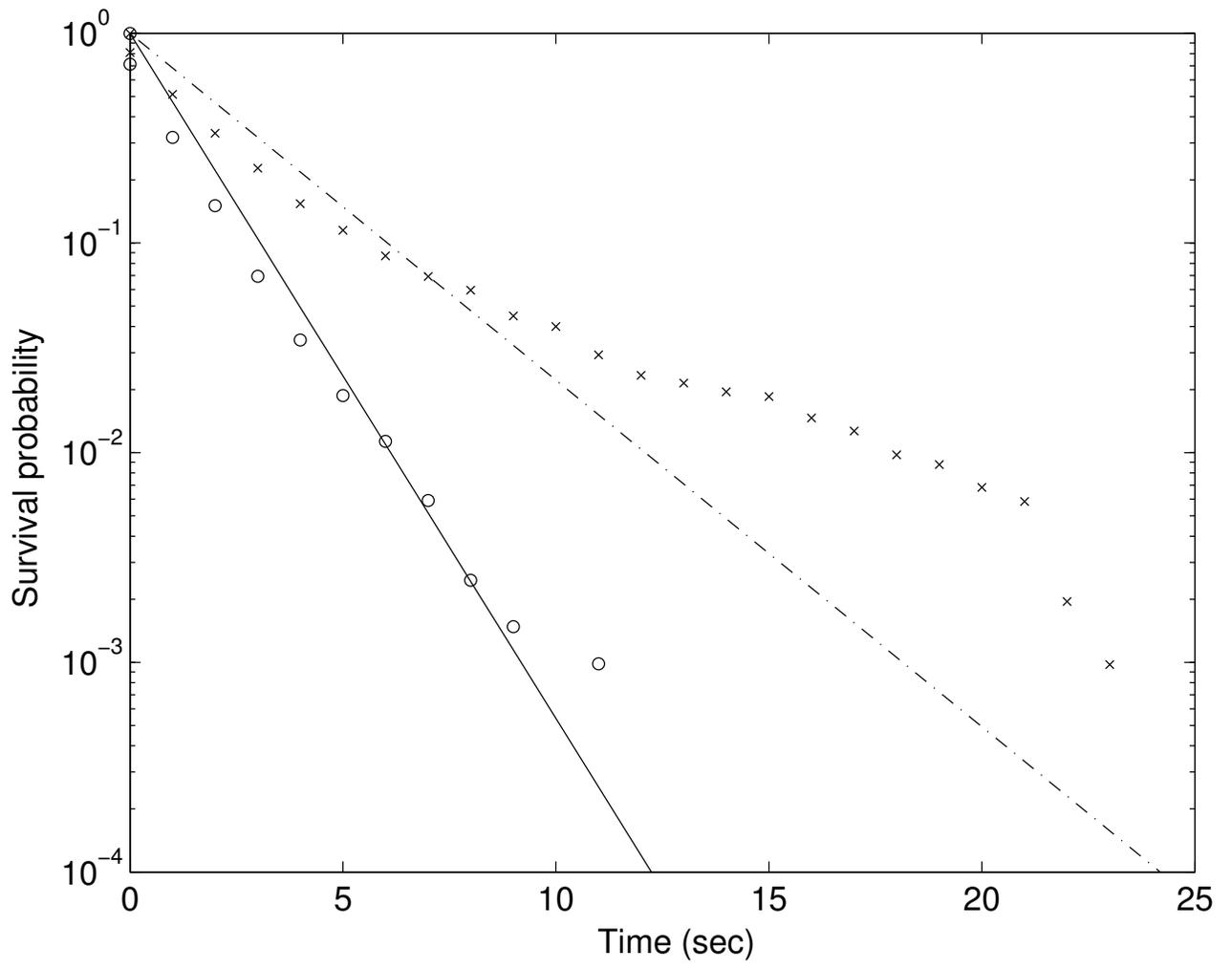

**Fig. 2f**

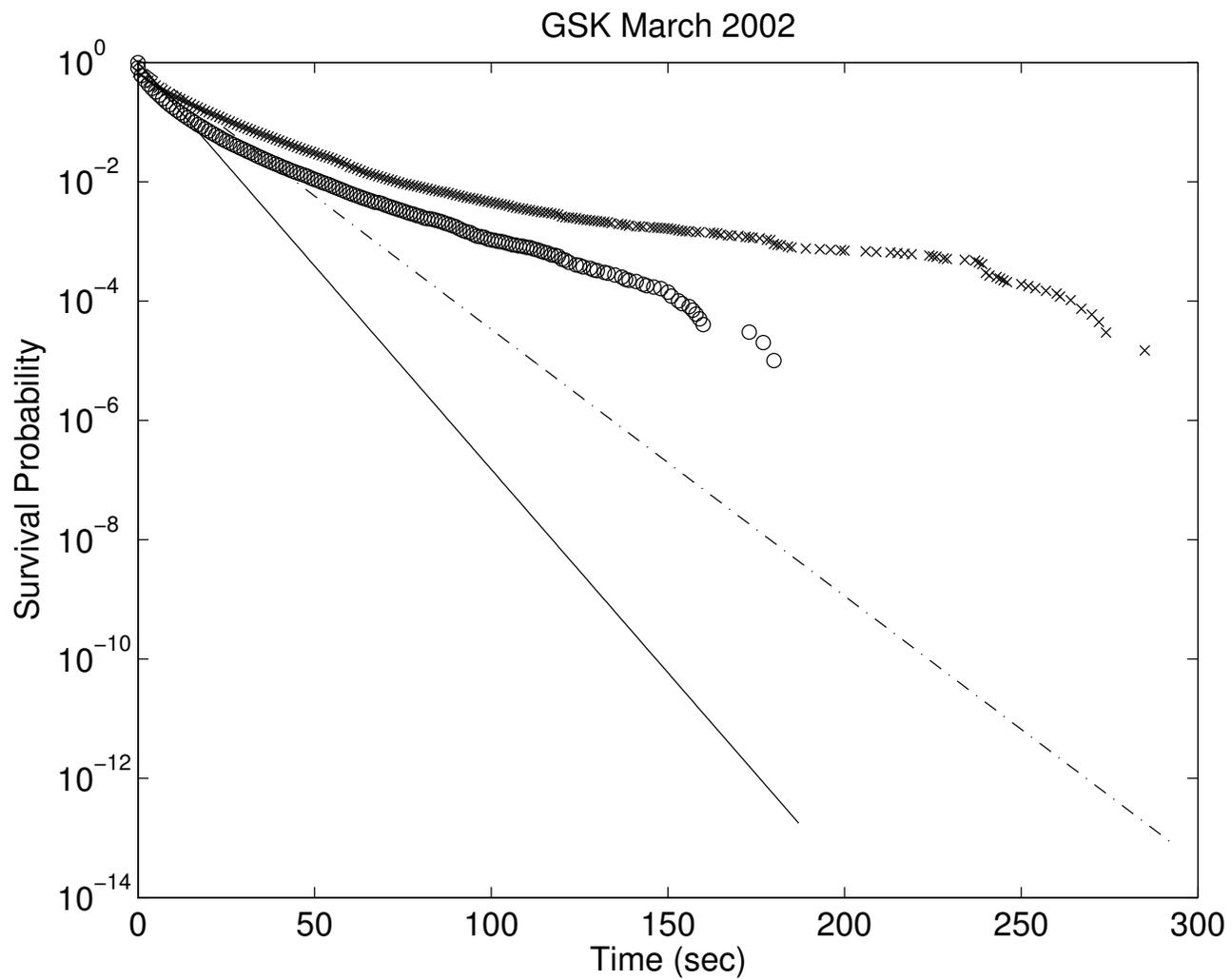

**Fig. 3**